\documentclass{article}
\usepackage{spconf,amsmath,graphicx,amssymb,enumitem,float}
\def\C{{\mathbb C}}
\DeclareMathOperator*{\argmin}{argmin\,}
\newcommand\norm[1]{\lVert#1\rVert}

\title{A Generative Diffusion Model to solve Inverse Problems for Robust in-NICU Neonatal MRI}
\name{Yamin Arefeen$^{\star \dagger}$ \qquad Brett Levac$^{\star}$ \qquad Jonathan I. Tamir$^{\star}$\thanks{Funding: Aspect Imaging, NSF CCF-2239687, IFML 2019844, JCCO}}
\address{$^{\star}$ The University of Texas at Austin \qquad $^{\dagger}$ MD Anderson Cancer Center}
\begin{document}
%
\maketitle
\begin{abstract}
We present the first acquisition-agnostic diffusion generative model for Magnetic Resonance Imaging (MRI) in the neonatal intensive care unit (NICU) to solve a range of inverse problems for shortening scan time and improving motion robustness. In-NICU MRI scanners leverage permanent magnets at lower field-strengths (i.e., below 1.5 Tesla) for non-invasive assessment of potential brain abnormalities during the critical phase of early live development, but suffer from long scan times and motion artifacts. 
In this setting, training data sizes are small and intrinsically suffer from low signal-to-noise ratio (SNR). This work trains a diffusion probabilistic generative model using such a real-world training dataset of clinical neonatal MRI by applying several novel signal processing and machine learning methods to handle the low SNR and low quantity of data. The model is then used as a statistical image prior to solve various inverse problems at inference time without requiring any retraining. Experiments demonstrate the generative model's utility for three real-world applications of neonatal MRI: accelerated reconstruction, motion correction, and super-resolution.
\end{abstract}
\begin{keywords}
Generative Models, Inverse Problems, MRI, Low-field, Neonatal
\end{keywords}
\section{Introduction}
\label{sec:intro}

Recently, machine learning algorithms for solving inverse problems in magnetic resonance imaging (MRI) yield state-of-the art results \cite{heckel2024ml}. End-to-end methods learn a point-wise mapping between measured data and reconstructed image, but are highly susceptible to test-time shifts in the measurement operator. On the other hand, generative methods that learn a prior over clean images have found application in MRI inverse problems as they are more robust to test-time distribution shifts \cite{jalal2021robust,chung2022score}. While powerful, both techniques require high quality and large quantity training data. 

MRI of infants using lower-field (below 1.5 tesla) scanners sited directly inside the neonatal intensive care (NICU) enables non-invasive assessment of potential brain abnormalities during the critical phase of neonatal and preterm development \cite{woodward2006neonate, inder2013neonate, dubois2020neonate}. However, many patients are unable to stay still during the lengthy scan times, so motion remains a challenge that precludes MRI access to many sick infants.

Motion in adult MRI scans are mitigated by solving inverse problems that reconstruct images from undersampled acquisitions with reduced scan times and by correcting motion corrupt data \cite{frost2022mocoreview}. Clinics accelerate MRI  by combining parallel imaging \cite{deshmane2012pi} -- which exploits the multi-channel signal receive array, hand-crafted spatial regularization, Compressed Sensing \cite{shreyas2007pcs}, and partial Fourier super-resolution \cite{mcgibney1993pf}. Recently, clinical practices and commercial products have also started adopting machine learning methods \cite{kiryu2023clinicaldl}.

The unique challenges of lower-field neonatal MRI precludes direct application of the aforementioned methods to improve motion robustness of adult MRI. Many lower-field systems use a single channel receive coil \cite{thiim2022neonate}, so parallel imaging cannot be applied. Machine learning algorithms using models trained on adult patients cannot be used because brain structure can vary greatly between neonates and adults \cite{fiagji2017neonate}. In addition, the images are inherently noisier in lower-field neonatal MRI and the permanent magnets used induce field inhomogeneity artifacts typically not seen with superconducting magnets in standard adult MRI. Finally, there are fewer publicly available data repositories, making it challenging to train machine learning models for inverse problems, from both a data quality and quantity perspective.

To the best of our knowledge, this work presents the first diffusion-probabilistic generative model trained on real-world, lower-field, in-NICU neonatal MRI data from a diverse range of image contrasts and anatomical orientations to solve various inverse problems that accelerate acquisitions and improve motion robustness. Our contributions are:
\begin{itemize}[left=0pt,nosep]
\item We establish a novel training dataset of real-world, lower-field clinical neonatal MR images acquired with the in-NICU 1T Tesla Embrace System (Aspect Imaging, Ltd) in collaboration with Sha'are Zedek Medical Center.
\item We apply a number of machine learning methods when training the generative model to address the challenging real-world nature of our dataset: \textbf{(1)} modifying existing popular diffusion network architectures to support inputs with varying matrix sizes, a common variability in MRI, therefore expanding the set of potential training images; \textbf{(2)} training a single model on combined data from all contrasts and orientations with class embeddings rather than stretching the dataset thin by training a separate model for each class; and \textbf{(3)} applying self-supervised denoising to boost the SNR of our dataset before training.
\item We experimentally demonstrate  the model's performance on real-world, motion-corrupt, clinical data from the NICU for three inverse problems: accelerated reconstruction, motion correction, and super-resolution.
\end{itemize}

The proposed method reduces scan time of single-coil Fast Spin Echo and Spin Echo acquisitions by an average of $1.5\times$ and  $2\times$ by using the generative model to reconstruct high-fidelity images from realistically undersampled measurements. Then, we use the acquisition agnostic property of our generative prior to solve motion correction inverse problems where the measurement model changes for each test example, reducing motion artifacts in prospectively acquired data. Finally, we show that shorter, lower-resolution acquisitions can be sharpened by solving a super-resolution inverse problem with the generative model.

\section{Theory}
\label{sec:theory}
\subsection{Accelerated Neonatal MRI with Diffusion Models}
Consider the single-channel MRI measurements $y = \mathbf{N}_K x + \eta$ in our neonatal setting, where $x \in \C^n$ is the vectorized clean image, $\mathbf{N}_K \in \C^{m\times n}$ is the 2D (possibly non-Cartesian) Fourier transform evaluated at coordinates $K$, $\eta \in \C^m$ is Gaussian random noise, and $y \in \C^m$ are the acquired Fourier measurements, or so called `k-space'. When the number of measurements matches the number of image pixels and SNR is high, solving the following inverse problem yields a suitable image:
\begin{equation}
    \argmin_x \norm{y-\mathbf{N}_K x}_2^2.
    \label{eq:yAx}
\end{equation}
MRI scan time can be reduced by acquiring fewer measurements than image pixels (i.e., $m<n$) through undersampling (accelerated MRI) or decreasing resolution (super-resolution) but this results in an ill-posed inverse problem that yields non-diagnostic images without suitable regularization. 

Generative models solve ill-posed inverse problems by learning the statistical prior, $p(x)$, over clean images to guide the reconstruction toward solutions that both match the data and are statistically likely. Specifically, diffusion models \cite{song2021sde, karras2022edm} indirectly learn $p(x)$ by training a neural network (NN) $D_\theta(x)$ to approximate the score $\nabla_{x_t} \log p_t(x_t)$ of progressively noised distributions. Then, the process of reconstructing an image can be viewed as approximating the conditional expectation $\mathbb{E}[x|y]$, i.e., taking the average of multiple samples from the posterior distribution $x \sim p(x|y)$. We sample from the posterior distribution using an Euler solver on the reverse ordinary differential equation (ODE) 

\begin{equation}
    dx = \left[\frac{\dot{s}(t)}{s(t)}x - s(t)^2 \dot{\sigma}(t)\sigma(t)\nabla_x \log p\left( \frac{x}{s(t)} | y; \sigma(t) \right)\right] dt
\end{equation} with $s(t)=1$ and $\sigma(t)=t$ following \cite{karras2022edm}. Using Bayes rule we can separate the posterior score into likelihood and prior scores,
\begin{equation}
    dx = \left[-t \left(\nabla_x \log p\left( y|x ; t \right) + \nabla_x \log p\left( x ; t \right)  \right)\right] dt.
\end{equation}
Substituting the pretrained diffusion model for the score function of the prior and the analytical expression for the likelihood score \cite{jalal2021robust, chung2023dps} yields the final ODE used for posterior sampling, 
\begin{equation}
    dx = \left[-t \left(\nabla_x \norm{\mathbf{N}_K\tilde{x}(x) - y}_2^2 + D_\theta(x,t\right) \right] dt
\end{equation}
The analytical expression for the likelihood score is technically only known at time point $t=0$ so we use the approximation $\tilde{x}(x) = \mathbb{E}[x_0|x]$ \cite{chung2023dps}. Note how this formulation decouples the statistical prior from the likelihood, so for neonatal MRI, a pre-trained prior can be re-used to solve inverse problems with different sampling patterns, receive coils, timings, and measurement models.

\subsection{Motion Correction}
Motion in MRI can be formulated as uncertainty in the measurement model. Let $\kappa = \{\phi, \theta\}$ be a variable which holds all information about the 2D rigid body motion, i.e., the translations and rotations that occurs during the scan. Then the measurement model is $A_\kappa = P_{\phi}N_{R_{\theta}K}$, where $R_{\theta}$ is a rotation matrix for all motion states, $P_{\phi}$ is a diagonal matrix implementing a linear phase shift describing the horizontal and vertical translations during each motion state, and $N_{R_{\theta}K}$ is the non-uniform Fast Fourier Transform (NUFFT) of $x$ at the coordinates $R_{\theta}K$. Then, one can estimate a clean image, and its associated motion parameters, from an acquisition in the presence of motion by solving, 
\begin{equation}
\label{eq:motion}
    \argmin_{x,\kappa} ||y- A_\kappa x||_2^2.
\end{equation}
Assuming a uniform prior on the motion, we can sample from the joint posterior $p(x,\kappa|y)$ by solving the reverse ODE with Euler updates, and for a more detail see \cite{levac2024moco}. 
 
The acquisition-agnostic property of diffusion models becomes particularly advantageous for this motion correction inverse problem. As a neonatal patient in the NICU experiences arbitrary and unpredictable motion, it is impractical to build a training dataset that captures all possible motion states that might be encountered at inference time. Because diffusion-based posterior sampling decouples the prior and likelihood, the diffusion model, $D_\theta$, trained on motion-free images can be applied to solve the motion inverse problem with arbitrary neonatal motion at test time.

\section{Methods}
\label{sec:methods}

\begin{figure}
\centering
\centerline{\includegraphics[width=.95\linewidth]{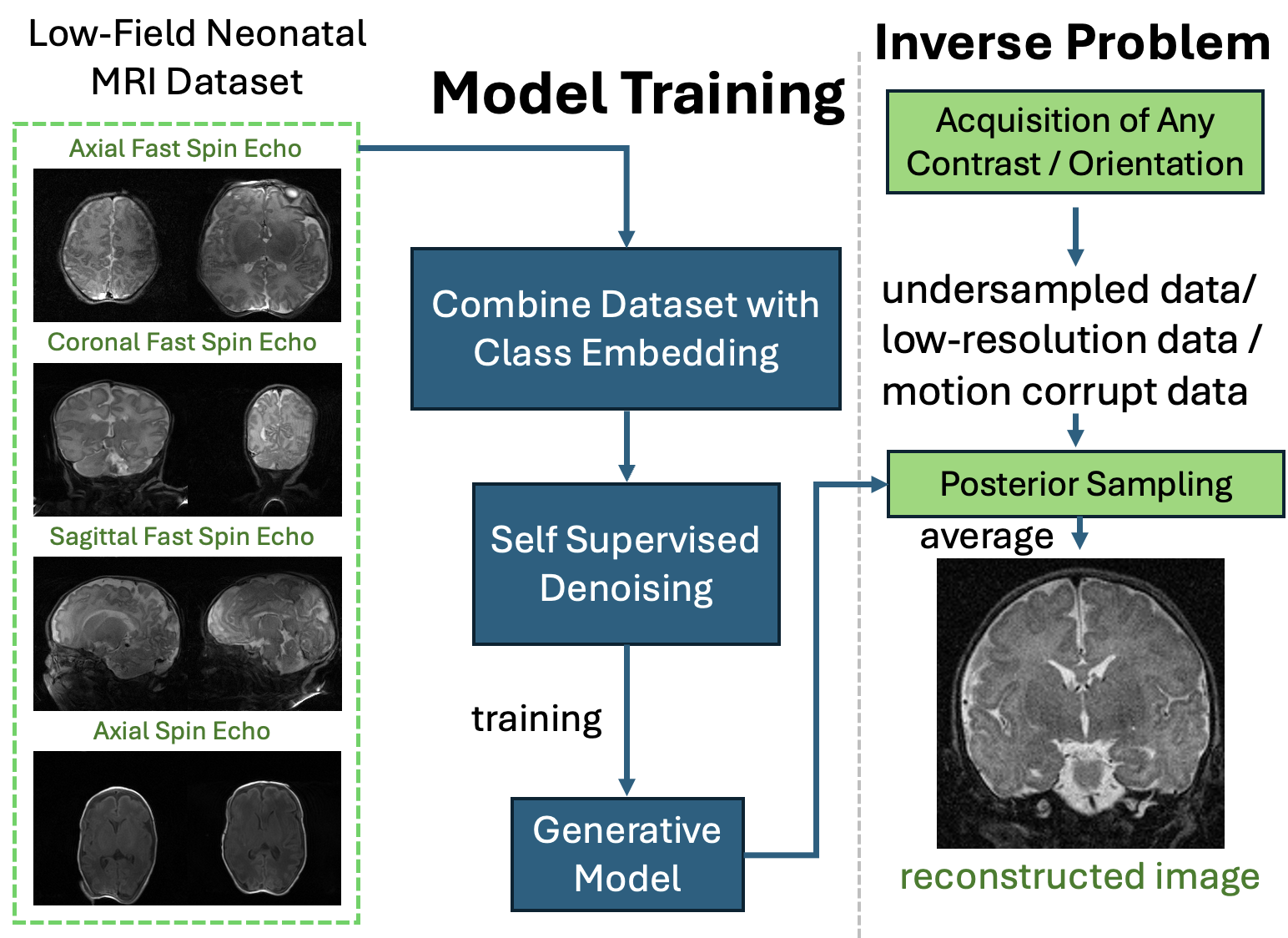}}
\caption{We propose training a generative model on low-SNR, low-quantity in-NICU neonatal data by combining all contrasts and orientations with class embedding and applying self-supervised denoising before training. The generative model applies to varies inverse problems to accelerate in-NICU neonatal MRI.} 
\label{fig:training}
\end{figure}

\subsection{Neonatal Dataset}
In collaboration with Aspect Imaging and Sha'are Zedek Medical Center under IRB approval and informed consent/assent, we gathered a dataset acquired with the in-NICU, 1 Tesla Embrace system from 128 neonatal subjects. Each subject was scanned with axial, coronal, and sagittal $T_2$ Fast Spin Echo (FSE) and axial $T_1$ Spin Echo (SE) sequences. We randomly split this dataset into 108 subjects for training and validation and 20 subjects for testing, resulting in $8659$ ($440$) FSE and $3224$ ($210$) SE training (testing)  slices. To simplify training, we re-sized, in k-space, all training slices to $200 \times 200$ matrix size. Field inhomogeneity correction available on the Embrace System was not applied to our dataset.

\subsection{Proposed Generative Model Training Pipeline}
Naive application of standard generative model training methods do not work well in the in-NICU neonatal MRI setting due to the low quantity of and corruptions in the data, therefore we developed a training pipeline to address these problems. We based our model on a U-Net \cite{song2021sde} style network with 65 million parameters and used the ``EDM'' hyper-parameters, loss function, and diffusion noise schedules \cite{karras2022edm}. 

To adapt training to our in-NICU setting, we first modified the down-sampling and up-sampling portions of the network architecture so that the model can take varying matrix size inputs. We set pre-determined resolutions at each U-Net level, based on the dimensions of the training data, and then perform the up-sampling or down-sampling with bi-linear interpolation to those resolutions. Thus, we can re-size and use all of our heterogeneous matrix size data for training and let the model handle matrix size discrepancies at inference. Second, we trained a single model on all the coronal, sagittal, and axial FSE and axial SE data simultaneously. Our method one-hot encodes each class and simultaneously trains a multi-layer perceptron that takes the encoding vector as input and outputs an embedding that serves as an additional input to each of the U-Net blocks in the larger model. In this way, the model uses image contrast and orientation information during training and inference. Finally, since we only have access to our inherently noisy dataset, we train another U-Net for denoising in a self-supervised fashion by applying the Noisier2Noise \cite{moran2020n2n} algorithm. This denoising model is applied to the training dataset before training our ``EDM'' based generative model. Figure \ref{fig:training} schematically illustrates our training procedure.

\subsection{Experiments}
We perform accelerated MRI reconstruction, motion correction, and super-resolution experiments to demonstrate the utility of our generative model for neonatal MRI across varying measurements models and inverse problems. 

\textit{Accelerated MRI reconstruction:} First, test axial (220), coronal (110), sagittal (110) FSE slices and axial (210) SE slices were retrospectively undersampled by an average rate of $2.0\times$ (corresponding to an equivalent reduction in scan time). We undersampled by throwing away groups of data associated with each echo train to achieve realistic undersampling with respect to signal decay \cite{rajput2024retro}. To analyze the effect of combining data with class embedding, we compared reconstruction performance of diffusion models trained on all data with and without class embeddings to diffusion models trained on each contrast and orientation separately. No method employed denoising pre-training in this experiment. Second, we under-sampled test the FSE and SE slices by $1.5\times$, and compared non-learned L1-wavelet \cite{shreyas2007pcs} based reconstructions to generative models trained with combined data using class embeddings with and without denoising.

\textit{Motion Correction:} We identified axial and coronal FSE acquisitions in the test dataset with motion artifacts and solved the inverse problem of Equation \ref{eq:motion} to estimate motion-free images and the associated motion parameters using the same proposed generative model applied in accelerated MRI reconstruction. We emphasize that this experiment used prospective clinical data, thus no ground truth exists, so we visually compare the original, motion-corrupt clinical image to our method.

\textit{Super-resolution:} We retrospectively lowered the resolution of example axial, coronal, sagittal FSE and axial SE slices by factor of $2.5 \times$ (corresponding to an equivalent reduction in scan time) by discarding the high-frequency k-space measurements. Super-resolution on this low-resolution data was performed by solving the inverse problem described in Equation \ref{eq:yAx} using the same proposed neonatal generative model applied in previous experiments.

\begin{figure*}[h!]
   \centering
   \centerline{\includegraphics[width=.9\linewidth]{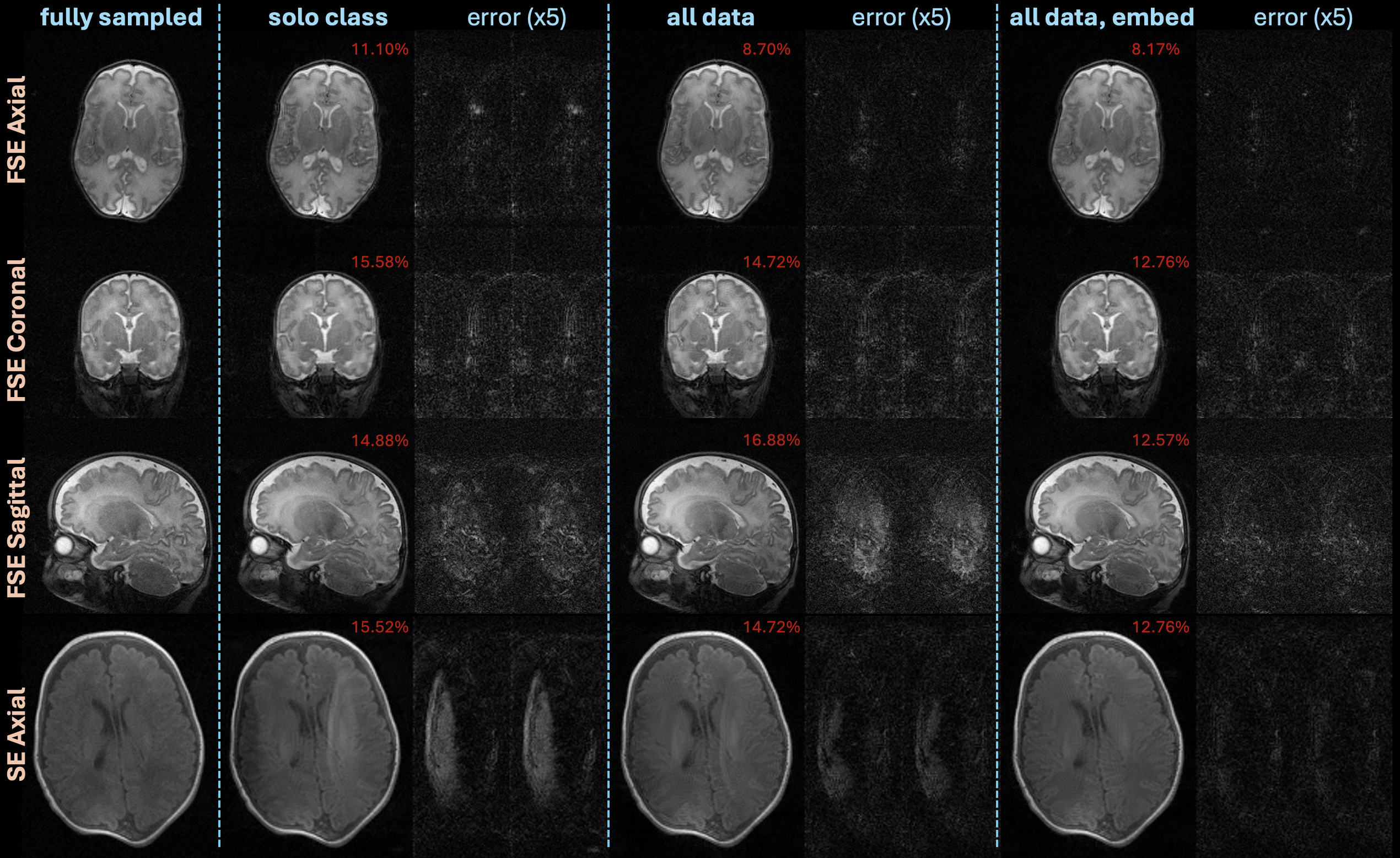}}
   \caption{Example accelerated ($2\times$ scan time reduction) MRI reconstructions using generative models trained on each contrast and orientation separately versus models trained on all data simultaneously with and without class embeddings. The proposed approach of training on all data with class embeddings achieves the best quantitative results and lower values in the error maps.}
   \label{fig:embed}
\end{figure*}

\section{Results}
\label{sec:results}
\subsection{Accelerated MRI Reconstruction}
Table \ref{tab:embed} presents normalized root mean squared error \break (NRMSE $\times 100$) comparisons of the $2\times$ accelerated reconstruction experiments when using generative models trained on each contrast and orientation separately versus models trained on all data with and without class embeddings. Figure \ref{fig:embed} visually compares example reconstructions for the various contrasts and orientations. Both models trained on all data achieved lower NRMSE in comparison to models trained separately. Additionally, incorporating embeddings improves performance when training using all data.

\begin{table}[h]
 \centering
 \caption{Accelerated MRI using all data and embeddings}
 \resizebox{\columnwidth}{!}{
 \begin{tabular}{|c|c|c|c|}
   \hline
    $(R=2.0)$& Solo & All Data & All Data, Embed\\  
   \hline
   Axial FSE & $13.30 \pm 2.97$ & $12.33 \pm 3.65$ & $\mathbf{11.95 \pm 3.44}$\\ 
   \hline
   Sagittal FSE & $16.76 \pm 2.77$ & $15.26 \pm 2.42$ & $\mathbf{15.01 \pm 2.63}$\\ 
   \hline
   Coronal FSE & $16.39 \pm 2.79$ & $14.36 \pm 2.64$ & $\mathbf{13.66 \pm 2.06}$\\ 
   \hline
   Axial SE & $12.46 \pm 2.05$ & $11.79 \pm 2.18$ & $\mathbf{11.56 \pm 2.38}$\\ 
   \hline
 \end{tabular}
 }
\label{tab:embed}
\end{table}

Table \ref{tab:denoising} compares quantitative performance of $1.5\times$ accelerated MRI reconstructed using the non-learned L1-wavelet method with generative models trained on all data using class embeddings with and without denoising pre-training. The learning based methods outperform L1-wavelet, and the proposed approach with denoising achieves comparable or superior average NRMSE in comparison to the proposed approach without denoising pre-training. Note, the quantitative results are complex-valued differences computed with respect to the fully-sampled, non denoised images \cite{haldarnoise}.

\begin{table}[h]
 \centering
 \caption{Accelerated MRI with and without denoising}
 \resizebox{\columnwidth}{!}{
 \begin{tabular}{|c|c|c|c|}
   \hline
    $(R=1.5)$& L1-wavelet & Ours No Denoising & Ours Denoising\\  
   \hline
   Axial FSE & $14.12 \pm 4.68$ & $9.03 \pm 2.16$ & $\mathbf{8.77 \pm 2.10}$\\ 
   \hline
   Sagittal FSE & $22.11 \pm 4.87$ & $12.31 \pm 2.08$ & $\mathbf{11.97 \pm 1.91}$\\ 
   \hline
   Coronal FSE & $21.60 \pm 10.56$ & $\mathbf{10.83 \pm 1.79}$ & $10.93 \pm 1.78$\\ 
   \hline
   Axial SE & $8.71 \pm 0.88$ & $8.96 \pm 1.00$ & $\mathbf{8.49 \pm 1.02}$\\ 
   \hline
 \end{tabular}
 }
\label{tab:denoising}
\end{table}
 
\subsection{Motion Correction}
\begin{figure}
   \centering
   \centerline{\includegraphics[width=.9\linewidth]{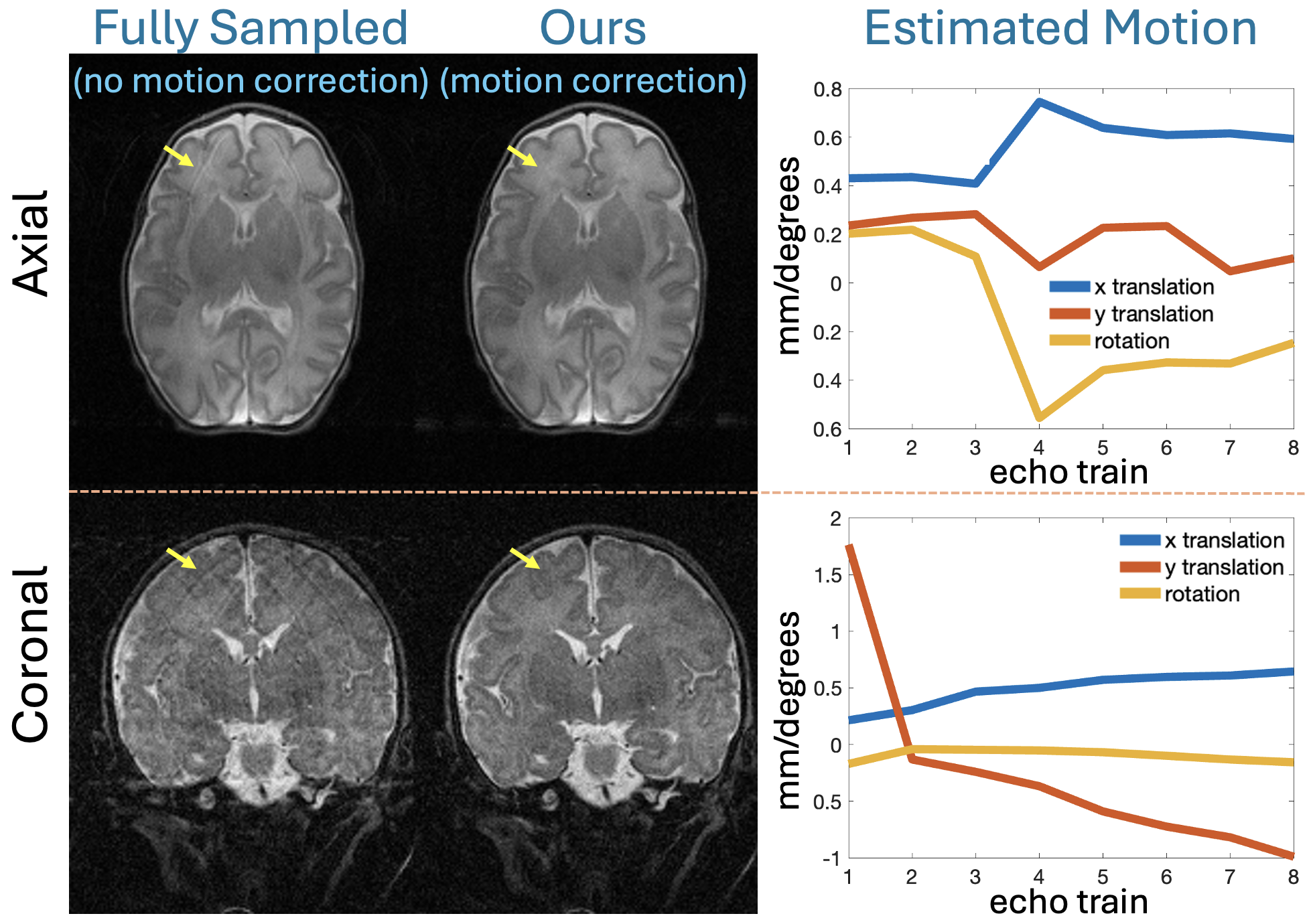}}
   \caption{Motion correction experiments on prospectively acquired clinical data in the presence of motion. The standard clinical image suffers from motion artifacts where our method uses the proposed generative model to reconstruct images with fewer artifacts and estimate the associated 2D rigid motion parameters.}
   \label{fig:motion}
 \end{figure}

Figure \ref{fig:motion} shows our generative model applied to the task of motion correction on prospectively measured, motion corrupt clinical data. The fully-sampled, clinical axial and coronal images suffer from artifacts induced by patient motion during the scan. Our motion correction method employs the same pre-trained generative model used for accelerated MRI reconstruction to simultaneously estimate a motion free image and associated motion parameters from motion corrupt data. Qualitative reduction in ringing artifacts, highlighted by the yellow arrows, is observed when using the proposed method.

\subsection{Super-resolution}
\begin{figure}
   \centering
   \centerline{\includegraphics[width=.9\linewidth]{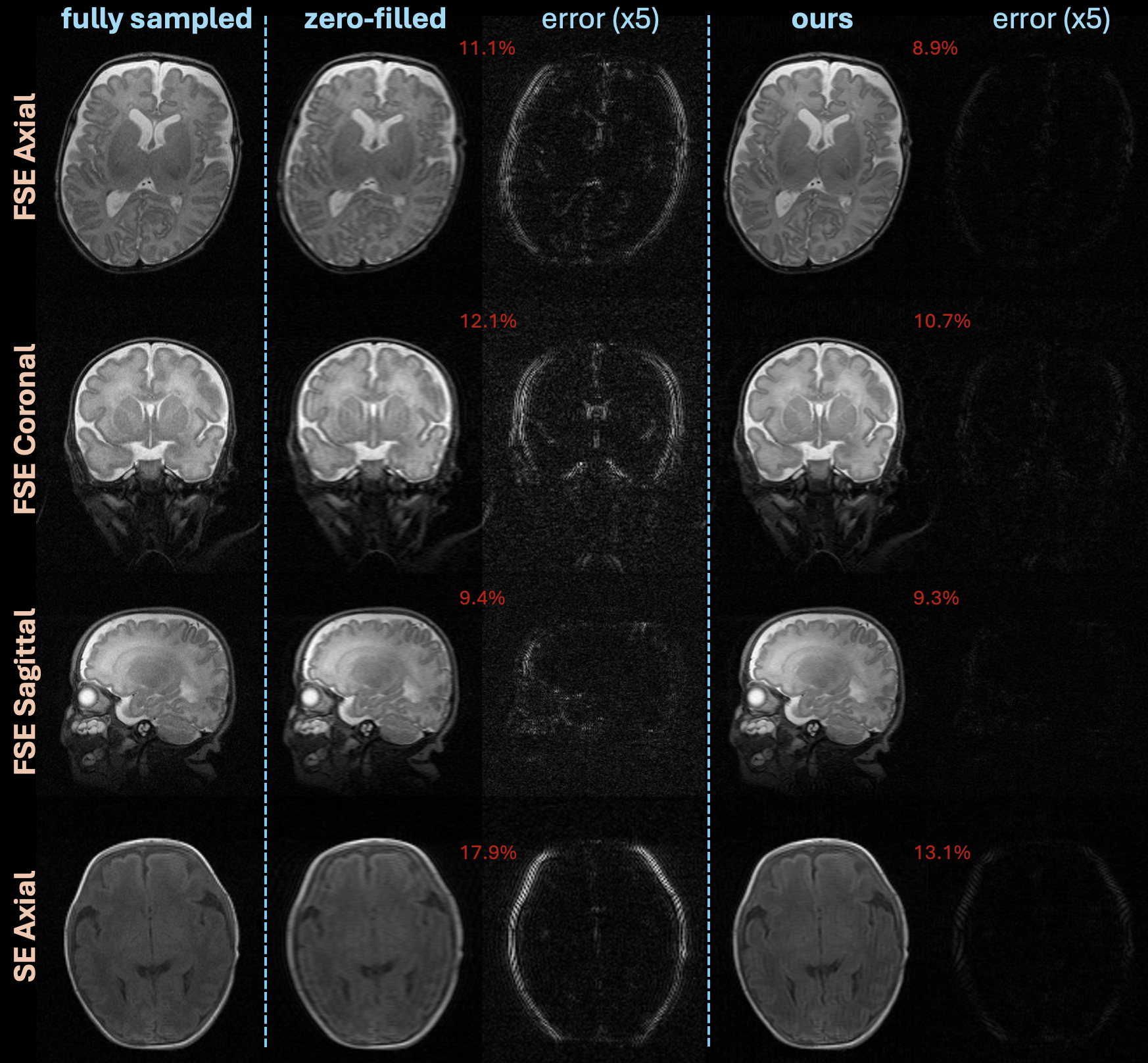}}
   \caption{Applying the proposed generative model to the super-resolution inverse problem. Limited extent in k-space sampling, corresponding to a $2.5\times$ reduction in scan time, results in low resolution images with Gibbs Ringing, but solving the super-resolution inverse problem with the proposed generative model improves image sharpness.}
   \label{fig:supres}
 \end{figure}

Figure \ref{fig:supres} demonstrates the generative model applied to solving the super-resolution inverse problem on axial, coronal, sagittal FSE and axial SE slices. Due to a the $2.5\times$ reduction in extent of k-space sampling, and therefore scan time, the images in the low-res column exhibit lower image resolution and Gibbs Ringing artifacts, but applying our generative model to solve the inverse problems results in sharper images.

\section{Discussion and Conclusions}
\label{sec:disc}
This study presented the first acquisition-agnostic diffusion probabilistic generative model for in-NICU lower-field neonatal MRI that serves as a prior to solve a variety of inverse problems to improve scan efficiency and motion robustness. We gathered a dataset of real-world clinical in-NICU images and developed a training pipeline to train a diffusion model on the low SNR and low quantity dataset. Then we apply the generative prior to the tasks of accelerated MRI reconstruction, motion correction, and super-resolution of axial, coronal, and sagittal FSE, and axial SE acquisitions. The decoupling of the diffusion prior and measurement model was particularly useful in the motion correction setting where a different measurement model is encountered for each test example.

While the proposed pipeline tackles noise corruption, our training dataset also consisted of images corrupted by motion artifacts. Future work will explore applying our generative model to identify motion corrupt images in the training dataset, correcting those images with the motion correction inverse problem, and re-training the generative model with the newly corrected images. In addition, the proposed pipeline addressed limited data availability by combining all data with class embeddings. Further improvements could be made by first pre-training the generative model on large, freely available adult MRI datasets, and then fine-tuning the model using our real-world, clinical in-NICU dataset \cite{hu2022lora}.

While this study presented initial quantitative and qualitative evaluation, clinical adoption requires further validation. First, this work presented proof-of-concept experiments showing that the generative model applies to the super-resolution setting, but evaluations on a larger test set with comparisons to appropriate baselines is needed to make quantitative performance conclusions. Second, a clinical reader study with board-certified radiologists who work with neonatal MR images to evaluate whether the accelerated and motion corrected images maintain diagnostic utility is needed. Finally, all inverse problems should be evaluated prospectively by acquiring undersampled data from the scanner in addition to retrospectively throwing away echo trains from fully-sampled data. We emphasize that our motion correction experiments in this work were prospective as we did not throw away any data before applying our algorithm.
 
\vfill\pagebreak

\end{document}